# Maximization of surface-enhanced transversal magneto-optic Kerr effect in Au/Co/Au thin films

**César Aurelio Herreño-Fierro**[*,1,2] **and Edgar J. Patiño**[**,1]

[1] Departamento de Física, Universidad de los Andes, Grupo de Física de la Materia Condensada, A.A. 4976-12340, Bogotá, Colombia
[2] Facultad de Ciencias y Educación, Universidad Distrital FJdC, Carrera 7 No. 40B - 53, Bogotá, Colombia

*c-herren@uniandes.edu.co, Phone: +57 1 3394949, Fax: +57 1 3324516
** Corresponding author: epatino@uniandes.edu.co, Phone: +57 1 3394949, Fax: +57 1 3324516

In order to maximize the transversal magneto optic Kerr effect (T-MOKE) of a Au/Co/Au structure we propose a method to obtain the optimum thickness values. A criteria based on preserving good plasmonic properties has been included as part of this method. Using the theoretical prediction we grew Au/Co/Au trilayers and perform optical and MO characterization using the Kretschmann configuration.

The results admit very easy interpretation in terms of the interaction between the magneto-optical and plasmonic properties dictating the optimal thicknesses of the structure. Moreover we have grown and characterized the optimized structure finding good agreement with theory reaching, for a 532 nm green laser, a maximal surface magneto-optic (MO) signal enhancement of close to nine folds with respect to the signal without plasmonic excitation.

## 1 Introduction

The interaction of electromagnetic field with electrons at the interface between a metal and a dielectric leads to a collective oscillation of conduction electrons known as surface plasmon polaritons (SPPs). These collective oscillations can be excited by incident light that matches its frequency and linear momentum. Furthermore it has been broadly demonstrated that SPPs enhance the magneto-optical activity of nano-structured systems [1-22]. This occurs even for those structures made of purely non-ferromagnetic metals under small magnetic fields [10]. This effect has been found to enhanced the sensing capabilities of biosensors [23, 24] and novel magnetoplasmonic devices [25, 26]. Such devices are currently a very active area of research. The aim is to achieve control of the SPP properties by external magnetic field [27, 28]. One way to achieve such control is to intercalate noble and ferromagnetic metals in a metal/ferromagnet/metal trilayer structure [5, 15, 16, 18, 19, 21]. In such structures called magnetoplasmonic (MP) materials, the noble-metal layers provide the plasmonic functionality given its low optical absorption and high quality factor; $|\mathrm{Re}\{\epsilon\}/\mathrm{Im}\{\epsilon\}| \gg 1$ where $\epsilon$ is the dielectric function, meanwhile the ferromagnetic layer provides the magneto-optical character.

Gold has been typically preferred as plasmonic material due to its very high chemical stability, i.e., resistant to corrosion and oxidation, although its optical absorption is not the lowest one. In Ref. [21] Ag has been used instead of Au given its lower optical absorption where the authors have found a larger enhancement in the MOKE signal. However for practical applications such structures have poor chemical stability. On the other hand cobalt has been typically preferred as MO material because of its high saturation magnetization offers a larger MO signal. Consequently the trilayer Au/Co/Au structure was chosen for this investigation.

It was previously investigated the effect of varying Co thickness of Au:5nm/Ag:5nm/Co:xnm/Ag:7nm structures on the enhancement of the magneto optical Kerr effect and modulation of the SPP wave vector [21]. In this study the Co thickness was varied between 0 and 11 nm. The authors found maximum enhancement of the magnetic response at a cobalt thickness of about 8nm. Also in Ref. [22] the magnetic modulation of the SPP wave vector was investigated as function of Au thickness of the top layer which corresponds to the separation between the Co layer surface and the interface sustaining SPP.

Motivated by the works of Ref. [21, 22] we focus our interest on the maximization of the MO response of Au/Co/Au where each of the layers thickness is varied based on a specific criteria; namely large MO response and low optical absorption. For this aim we propose a method in order to obtain the optimum structural parameters based on the scattering matrix method. Based on this theoretical prediction we grew Au/Co/Au trilayers and perform optical and MO characterization using the Kretschmann configuration for our experiments.

## 2 Theoretical methods

**2.1 Transversal magneto-optical Kerr effect and plasmonic excitation** The system under consideration is depicted in Fig. 1. The heterostructure glass/Au/Co/Au/air is illuminated, from the glass side, at an incident angle θ by a p-polarized coherent light source of 532 nm wavelength. In total reflection condition SPPs are excited at the Au/air interface (Kretschmann configuration [30]). The transversal MO Kerr effect (T-MOKE) configuration was chosen for the characterization given that it provides the largest MO enhancement [18]. Here the Co layer magnetization is in the plane of the sample (y axis), controlled by an external alternating magnetic field (B) along the same direction.

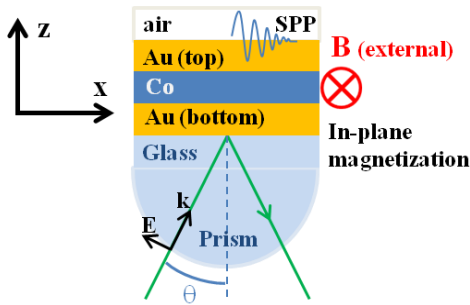

**Figure 1** Scheme of the system under consideration.

In this case the bulk dielectric tensor of the Co layer;

$$\epsilon_{Co} = \begin{pmatrix} \epsilon_x & 0 & M_y \epsilon_{xz} \\ 0 & \epsilon_x & 0 \\ -M_y \epsilon_{xz} & 0 & \epsilon_x \end{pmatrix} \quad (1)$$

is governed by the magnetization ($M_y$) [31]. The complex diagonal terms $\epsilon_x$ account for the isotropic response of the system while the off-diagonal anti-symmetric terms are identified as the MO constants. The external magnetic field in the *y* direction changes the magnetization of the ferromagnet between opposite saturation states.

In transversal geometry the MO Kerr effect consists on a change in the reflected intensity as function of magnetization. Therefore, the T-MOKE signal is defined as the reflectivity difference between the two opposite saturated magnetization states of the Co layer normalized by the reflectivity in demagnetized state R(0);

$$\frac{\Delta R}{R} = \frac{R(+M_y^s) - R(-M_y^s)}{R(0)} \quad (2)$$

where $R(\pm M_y^s)$ are the reflectivity for opposite magnetizations. Nevertheless, to avoid the artificial enhancement as result of the reflectivity minimum close to plasmon resonance angles, we define the T-MOKE signal as the change in reflectivity under magnetization inversion:

$$\Delta R = R(+M_y^s) - R(-M_y^s) \quad (3)$$

**2.2 Scattering matrix method** In order to model the optical and magneto-optical responses of noble-metal/ferromagnet/noble-metal structures, one has to take into account the presence of off-diagonal elements in the ferromagnet dielectric tensor. This fact does not allow a simple exact analytical solution. The first theoretical approach to investigate this was an analytical approximation based on the first Born approximation, considers the presence of the ferromagnet as a perturbation in a continuous structure [18, 19, 32, 33]. These are first order approximations that preserve acceptable accuracy for the case of ultrathin ferromagnetic layers compared to the total thickness of the plasmonic trilayer structure. Since in our study we consider the Co layer thickness to be comparable to the total thickness of the structure an exact numerical solution is preferred. The numerical approach is based on the scattering matrix method developed by [13, 29]. The general method is built up for periodically patterned multilayer systems. Nevertheless the formalism can be simplified for the case of uniform (non-structured) layers as early reported by Caballero and colleagues in reference [29].

**2.3 Parameters optimization** The set of optimal structural parameters (i.e., layers thickness) which maximize the enhancement of MO response of the system is obtained as follows. The optical constants for Au and Co used in this computation were obtained from [21]. The refractive index of the Prism/Glass-substrate system was experimentally measured by Brewster-angle experiment giving a value of 1.448, while the air refractive index was taken as 1.

We first chose an initial arbitrary set of parameters based on qualitative criteria. The initial structure was Au(30)/Co(4)/Au(0.5) where the numbers in parenthesis indicate thickness in nanometers. Here the total thickness is close to the optimal critical thickness for pure gold with maximum SP resonance at incident light of 532 nm wavelength. A 4 nm Co layer thickness was chosen as starting value to ensure the magnetization lies in plane and reduced optical absorption. The top Au layer is 0.5 nm of the order of Co layer roughness (see section 3). The same order of magnitude of roughness was obtained for the top Au layer which in principle led us to presume the Co layer to be fully covered by an ultra-thin gold layer. This permits the Co to be as close as possible to the air/Au interface where the SPP are sustained. Here the magnetic layer is exposed to a strong field at resonance producing the MO enhancement.

The next step is to optimize the structure. The process consists on calculating the reflectivity and the MO signal (i.e. T-MOKE) values in the angle range between 0 to 75 degrees for different layers thicknesses. In each step of the process two layer thicknesses are kept fixed and the remaining layer thickness is varied from 0 to 30 nm in steps of 0.1 nm. This process is repeated three times, one for each layer of the structure, in the following way.

First the two Au layers thicknesses are kept fixed and the Co layer thickness is varied. The results, which illustrate the general evolution of reflectivity and the MO signal, for six different Co thicknesses are shown in Fig. 2(a) and 2(b) respectively. From this, we obtain the optimal Co layer thickness equal to 10.2 nm for which the maximum T-MOKE signal is reached for the given Au layer thicknesses.

In the second computation, we fix the Co layer thickness to 10.2 nm, leave the Au top layer thickness equal to 0.5 nm, and vary the Au bottom layer thickness. The results for reflectivity and T-MOKE signal are shown in Fig. 2(c) and 2(d), respectively. Seven different curves illustrate Au bottom thickness general evolution. From this, we obtain the optimal Au bottom layer thickness to be 14.1 nm.

Finally, by fixing the Co and Au bottom layer thicknesses, we compute the last optimization for the Au top layer (Fig. 2(e) and 2(f)). In this case, as expected, the optimal value was found to be close to zero.

Figure 3 shows the evolution of the maximum T-MOKE signal as a function of each layer thickness. The symbols beside the curves correspond to experimental results to be discussed in section 3. As expected for the top Au layer the maximum T-MOKE signal decreases monotonically with thickness (red-dotted line). This can be easily understood as this layer determines the separation between the MO active Co layer and the air/Au-top-layer interface where the SPPs are excited. This separation prescribes the electric field intensity experienced by the Co layer.

On the other hand the bottom Au layer shows a non-monotonic behaviour symmetric around the maximal value of 14.1 nm corresponding to the optimal thickness (dashed blue line Fig. 3). This behaviour can be explained as the result of plasmon resonances that strongly depends non-monotonically on the thickness. As the thickness increases the quality factor and amplitude of the plasmon resonance improves and the reflectivity curves gets narrower and sharper until reaching the minimum reflectivity value at the critical thickness. When the sample thickness gets larger than the critical thickness, the resonance amplitude reduces. The previous explanation is consistent with data depicted on Fig. 3 (dashed blue line) corresponding to the bottom Au layer.

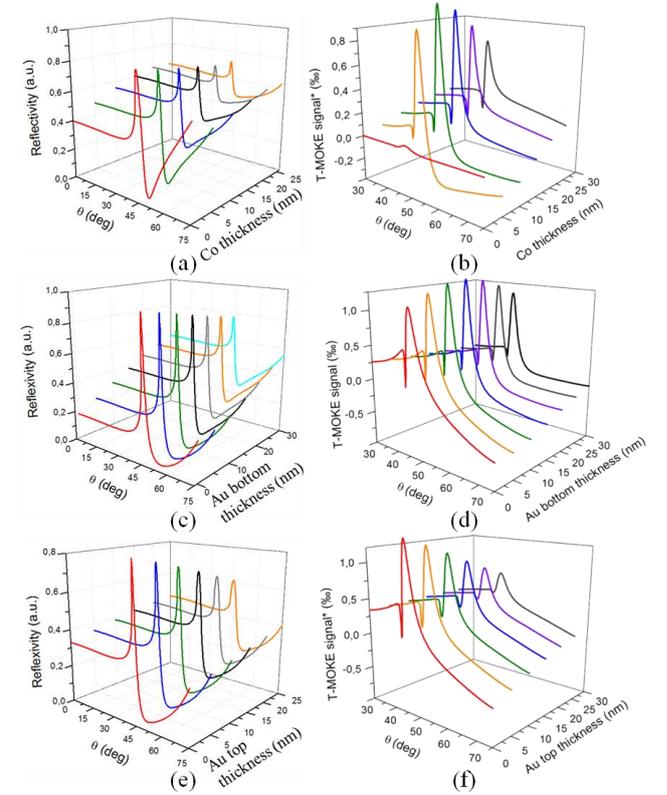

**Figure 2** Angular spectral reflectivity (a, c, and e) and T-MOKE signal (b, d, and f) as function of the thicknesses of; Co layer (a,b), Au bottom layer (c,d) and Au top layer (e,f). Each curve represents one thickness value.

Finally the maximal T-MOKE signal as function of Co thickness also shows a non-monotonic behaviour with a maximal value for a critical thickness of 10.2 nm in the Co layer (solid black line in Fig. 3). Note that this curve in not symmetrical around this value. This behaviour can be qualitatively explained as a competition of two effects. Firstly the T-MOKE signal is proportional to the Co layer thickness. Secondly an increase of the Co thickness should also led to a non-monotonic (increase/decrease) of the plasmon resonance quality. At first these effects cooperate, as shown by a pronounce slope of curve, improving the plasmon resonance. Later these effects compete showing a reduction of the T-MOKE maximum signal with a lower slope. Consequently these two effects are responsible of a non symmetrical shape of maximal T-MOKE signal.

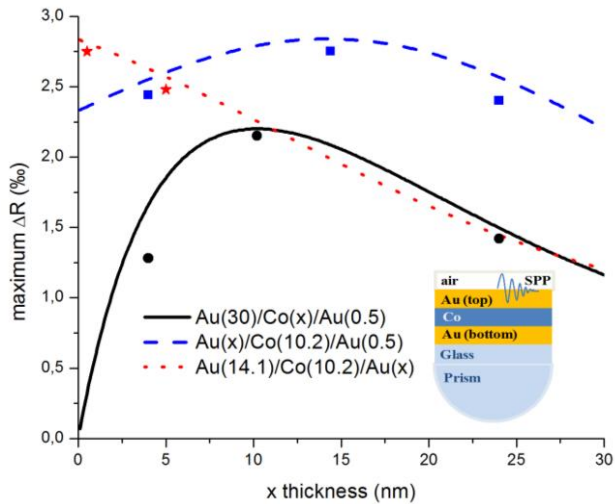

**Figure 3** Evolution of the maximum ΔR vs. thickness for each layer keeping the other two layers thicknesses fixed. Dependence on the Co layer thickness (solid black line), Au bottom layer thickness (dashed blue line), and the Au top layer thickness (dotted red line). Added symbols correspond to the experimental measurement of the maximum ΔR.

At the end of this process, the predicted optimized structure is the bilayer Au(14.1)/Co(10.2). Nevertheless to prevent the Co layer oxidation we include 0.5 nm Au layer on top of the Co giving the structure Au(14.1)/Co(10.2)/Au(0.5). This minimal Au top layer thickness ensures homogeneity condition as dictated by roughness measurements as detailed in section 3.1.
It should be pointed out that if a new iteration is done leaving the Co thickness as a free parameter a value of 16.9 nm is obtained however the optical absorption compromises the quality of the plasmonic resonance and therefore we kept the Co thickness as 10.2 nm as the optimal thickness of our structure.

## 3 Experimental methods
### 3.1 Sample growth and characterization
The optimized structure Au(14.1)/Co(10)/Au(0.5), and six additional structures (detailed in table 1) were grown and characterized in order to contrast the theoretical prediction with the experiment (symbols in Fig. 3).

The Au/Co/Au structures were grown using a UHV e-gun evaporation system. The base pressure was better than $3 \times 10^{-8}$ torr, and the evaporation pressure was less than $1.3 \times 10^{-7}$ torr. Each layers thickness was monitored during growth to better than 1 Å by quartz balance. The structures were grown on microscope cover slips (glass). Under these conditions the Co layer had in plane magnetic anisotropy and coercive fields of about 50 Oe (inset in Fig. 6(b)) as measured using a vibrating sample magnetometer (VSM). By atomic force microscopy (AFM) the RMS roughness was measured at the top of each of the layers obtaining the following values; 1.15 nm for glass, 1.19 nm for glass/Au, 0.47 nm for glass/Au/Co, and finally 0.57 nm for glass/Au/Co/Au. The fact the last to surfaces have similar RMS roughness demonstrates good uniformity of the Au top layer.

### 3.2 Experimental setup
The optical and MO characterization was made by angular spectral reflectivity and T-MOKE signal, respectively. A scheme of the magneto optics experimental setup is shown in Fig. 4. Here the T-MOKE signal is measured using a simple modulation technique. The sample is illuminated by a 533 nm p-polarized green laser trough a cylindrical lens. The reflected intensity, sensed by a photodiode is sent to a lock-in amplifier. A coil is set to apply a field along the sample plane normal to the plane of incidence (inset in Fig. 4). The current through the coil is varied by a square signal with a frequency of 3 Hz. This way the sample experiences a periodic magnetization.

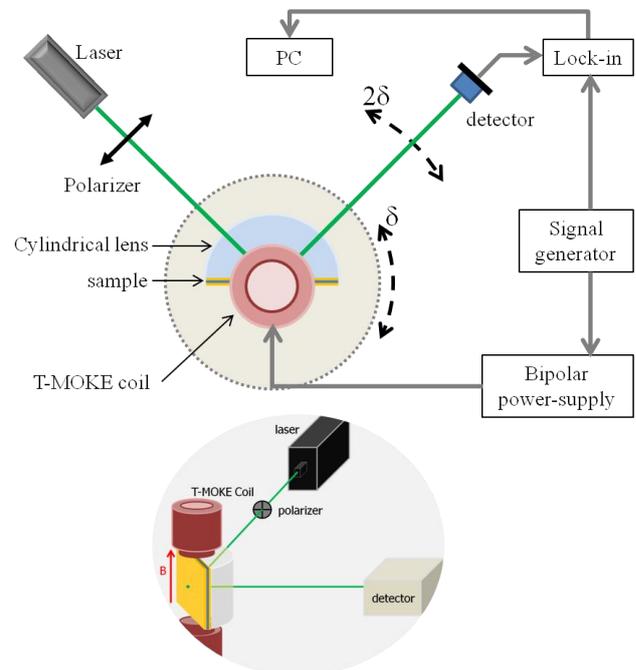

**Figure 4** Scheme of the magneto optics experimental setup. The inset shows a 3D scheme of the transversal MOKE geometry in the setup.

The current signal serves as a reference for the lock-in amplifier while measuring the photodiode voltage. Only the photodiode voltage in phase with the field is filtered by the locking amplifier allowing measurements of a signal to noise ratio better than 1/1000.

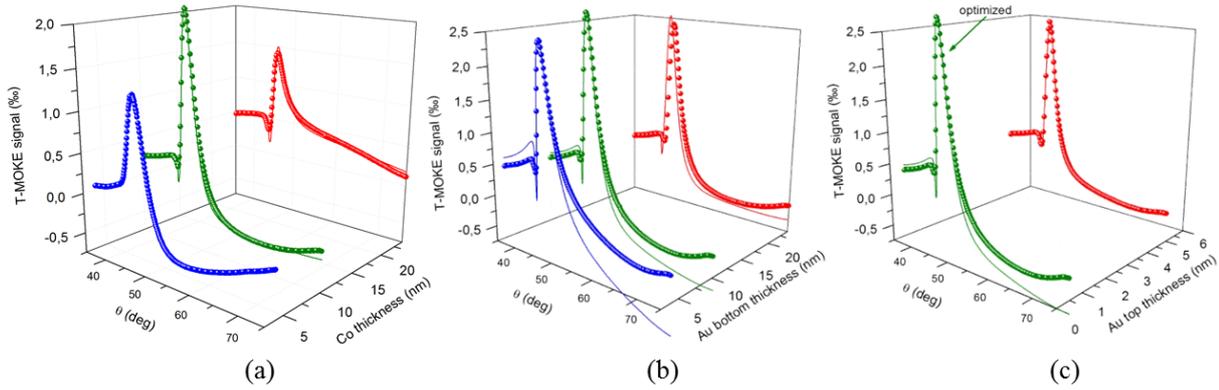

**Figure 5** Experimental results (spheres) and theoretical prediction (solid lines) for the optimization process of the Co layer (a), Au bottom layer (b), and Au top layer (c).

**3.3 Results and interpretation** The obtained experimental results are illustrated in Fig. 5 and summarized in table 1. In Fig. 5 the experimental measurements (solid spheres) and theoretical predictions (lines) of the T-MOKE signal are given for the first (a), second (b) and third (c) optimization processes. These results match fairly well the theoretical prediction showing the largest T-MOKE signal in all cases for varying thickness of Co, Au bottom and top layer respectively represented in green color (online version).

In Fig. 6 we show the optical (a), MO (b), and magnetic (inset) characterization for the optimized structure. As it can clearly be seen, the experimental results (solid spheres) show good agreement with the theoretical prediction (solid lines). The maximum MO signal experimentally obtained for this structure is of 2.75 per mil of the incident intensity. This value corresponds to about 9 times the MOKE signal in the absence of plasmon excitation (below the total reflection angle of incidence).

It is worth pointing out that in this work the reported enhancement of the MOKE has been measured compared to incidence intensity excluding the artificial values that may be obtained when normalizing with reflectivity.

**Table 1** Summary of experimental results. These points are plotted in Fig. 3. Co thickness

| Sample Au/Co/Au* | $\Delta R_{max}$ (‰) |
|---|---|
| Au(30)/Co(4)/Au(0.5) | 1,28 |
| Au(30)/Co(10.2)/Au(0.5) | 2,15 |
| Au(30)/Co(24)/Au(0.5) | 1,42 |
| Au(4)/Co(10.2)/Au(0.5) | 2,44 |
| **Au(14.1)/Co(10.2)/Au(0.5)** | **2,75** |
| Au(24)/Co(10.2)/Au(0.5) | 2,40 |
| Au(14.1)/Co(10.2)/Au(5) | 2,48 |

* Thicknesses expressed in nanometers (nm).

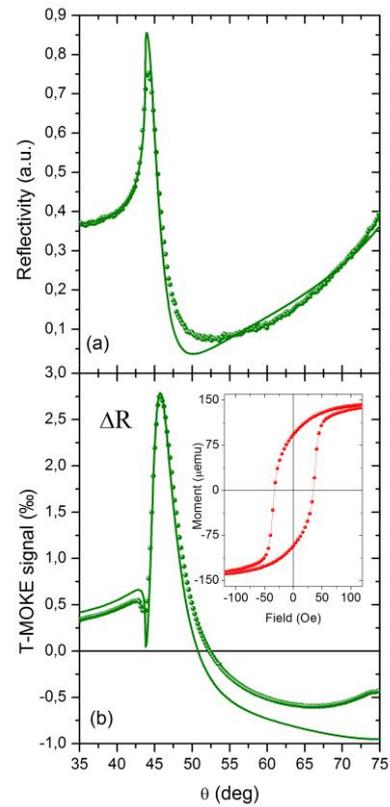

**Figure 6** Optical (a) and MO (b) characterization of the optimized sample. The inset shows the magnetic hysteresis loop obtained by VSM.

Finally a comparison between the experimental data (symbols) and theoretical curves (lines) in Fig. 3 shows the same trend and good agreement with each other.

## 5 Conclusions

We have implemented numerical calculations to compute the optimal structural parameters of a trilayer Au/Co/Au system which maximizes the T-MOKE signal. The resulting evolution profiles of the maximum MO signal admit very easy interpretation in terms of the interaction between the magneto-optical and plasmonic properties of the structure. Moreover, we have grown and characterized the optimized and other six structures finding good agreement with theory. This procedure can be applied to obtain the optimized structures when illuminated by longer wavelength light sources for which the surface enhancement is higher.

Here the Au-bottom and Co curves are non-monotonic while the Au top curve is monotonic with layer thickness. The non-symmetrical shape of the Co Curve is explained as the result of a cooperation (larger slope) and competition (smaller slope) between the MOKE and plasmon resonance effects.

**Acknowledgements** We thank E. Moncada-Villa from Departamento de Física, Universidad del Valle (Cali, Colombia) for providing the numerical data and useful discussions. We also thank A. García-Martín from Instituto de Microelectrónica de Madrid – Centro Nacional de Microelectrónica (Madrid, Spain) for providing the optical constants for the computation and very enlightening discussions. We acknowledge technical support from Luis Gómez-Chimbi in the design and construction of our experimental setup. This work was supported by "Programa Nacional de Ciencias Básicas" COLCIENCIAS (No. 120452128168), Vicerrectoría de Investigaciones and Facultad de Ciencias of Universidad de los Andes (Bogotá, Colombia), as well as Universidad Distrital (Bogotá, Colombia) for financial support.